\documentstyle[10pt,epsf,epsfig,
dp_delphititle,oldlfont,lineno,amssymb]{dp_delphi}
\makeindex
\def\DpPaperGroup{EP}
\def\DpPaperRef{2003-027}
\def\DpDate{1 Avril 2003}
\def\DpAuthors{DELPHI Collaboration}
\def\DpTitle{{ Measurement of Inclusive ${\bf f_1}$(1285) 
and ${\bf f_1}$(1420) Production in Z Decays with the DELPHI Detector}}
\def\DpSubmit{(Phys. Lett. B569 (2003) 129-139)}
\def\DpComment{  }
\def\DpEMail{   }
\def\jpsi{\hbox{$J{\kern-0.24em}/{\kern-0.14em}\psi$}}
\def\ev#1#2{\hbox{#1e{\kern-0.10em}V{\kern-0.30em}/{\kern-0.14em}$#2$}}

\def\gevcc{\ev{G}{c^2}}
\def\mevc{\ev{M}{c}}
\def\mevcc{\ev{M}{c^2}}
\def\tbhead#1#2{Table\ #1.\ #2\vskip8pt}

\catcode`@=11
\def\eqalign#1{\null\,\vcenter{\openup\jot\m@th
\ialign{\strut\hfil$\displaystyle{##}$&$\displaystyle{{}##}$\hfil
\crcr#1\crcr}}\,}
\catcode`@=12
\def\bqt#1#2\eqt{\begin{equation}\label{#1}%
{#2}\end{equation}\noindent}
\def\bln#1#2\eln{\begin{equation}\label{#1}%
\eqalign{#2}\end{equation}\noindent}
\def\brlist{\begin{thebibliography}{99}}
\def\brf{\bibitem}
\def\erlist{\end{thebibliography}}
\def\bra{\langle}
\def\ket{\rangle}
\begin{document}
\makeatletter
\newcount\@tempcntc
\def\@citex[#1]#2{\if@filesw\immediate\write\@auxout{\string\citation{#2}}\fi
  \@tempcnta\z@\@tempcntb\m@ne\def\@citea{}\@cite{\@for\@citeb:=#2\do
    {\@ifundefined
       {b@\@citeb}{\@citeo\@tempcntb\m@ne\@citea\def\@citea{,}{\bf ?}\@warning
       {Citation `\@citeb' on page \thepage \space undefined}}%
    {\setbox\z@\hbox{\global\@tempcntc0\csname b@\@citeb\endcsname\relax}%
     \ifnum\@tempcntc=\z@ \@citeo\@tempcntb\m@ne
       \@citea\def\@citea{,}\hbox{\csname b@\@citeb\endcsname}%
     \else
      \advance\@tempcntb\@ne
      \ifnum\@tempcntb=\@tempcntc
      \else\advance\@tempcntb\m@ne\@citeo
      \@tempcnta\@tempcntc\@tempcntb\@tempcntc\fi\fi}}\@citeo}{#1}}
\def\@citeo{\ifnum\@tempcnta>\@tempcntb\else\@citea\def\@citea{,}%
  \ifnum\@tempcnta=\@tempcntb\the\@tempcnta\else
   {\advance\@tempcnta\@ne\ifnum\@tempcnta=\@tempcntb \else \def\@citea{--}\fi
    \advance\@tempcnta\m@ne\the\@tempcnta\@citea\the\@tempcntb}\fi\fi}
 
\makeatother
\begin{titlepage}
\pagenumbering{roman}
\CERNpreprint{\DpPaperGroup}{\DpPaperRef} 
\date{{\small\DpDate}} 
\title{\DpTitle} 
\address{\DpAuthors} 
\begin{shortabs} 
\noindent
DELPHI results are presented on the inclusive 
production of two $(K\bar K\pi)^0$ states in the mass region
1.2--1.6 \gevcc\ in hadronic $Z$ decays at LEP I. 
The measured masses (widths) are $1274\pm6$ \mevcc\  ($29\pm12$ \mevcc) 
and $1426\pm6$ \mevcc\  ($51\pm14$ \mevcc) respectively.  
A partial-wave analysis of the $(K\bar K\pi)^0$ system shows that
the first peak is consistent with the 
$I^G(J^{PC})=0^+(1^{++})/(0^{-+})\,a_0(980)\pi$ and the second 
with the $I^G(J^{PC})=0^+(1^{++})\,K^*(892)\bar K+c.c.$ assignments. 
The  total hadronic production rates per hadronic $Z$ decay are 
$(0.165\pm0.051)$ and $(0.056\pm0.012)$ respectively.
These measurements are consistent with the two states being the $f_1(1285)$ 
and $f_1(1420)$ mesons.

\end{shortabs}
\vfill
\begin{center}
\DpSubmit \ \\ 
\DpComment \ \\
\DpEMail \ \\
\end{center}
\vfill
\clearpage
\headsep 10.0pt
\addtolength{\textheight}{10mm}
\addtolength{\footskip}{-5mm}
\begingroup
%
\newcommand{\DpName}[2]{\hbox{#1$^{\ref{#2}}$},\hfill}
\newcommand{\DpNameTwo}[3]{\hbox{#1$^{\ref{#2},\ref{#3}}$},\hfill}
\newcommand{\DpNameThree}[4]{\hbox{#1$^{\ref{#2},\ref{#3},\ref{#4}}$},\hfill}
\newskip\Bigfill \Bigfill = 0pt plus 1000fill
\newcommand{\DpNameLast}[2]{\hbox{#1$^{\ref{#2}}$}\hspace{\Bigfill}}
%
\footnotesize
\noindent
\DpName{J.Abdallah}{LPNHE}
\DpName{P.Abreu}{LIP}
\DpName{W.Adam}{VIENNA}
\DpName{P.Adzic}{DEMOKRITOS}
\DpName{T.Albrecht}{KARLSRUHE}
\DpName{T.Alderweireld}{AIM}
\DpName{R.Alemany-Fernandez}{CERN}
\DpName{T.Allmendinger}{KARLSRUHE}
\DpName{P.P.Allport}{LIVERPOOL}
\DpName{U.Amaldi}{MILANO2}
% ...
\normalsize
\endgroup
\titlefoot{Department of Physics and Astronomy, Iowa State
     University, Ames IA 50011-3160, USA
    \label{AMES}}
\titlefoot{Physics Department, Universiteit Antwerpen,
     Universiteitsplein 1, B-2610 Antwerpen, Belgium \\
     \indent~~and IIHE, ULB-VUB,
     Pleinlaan 2, B-1050 Brussels, Belgium \\
     \indent~~and Facult\'e des Sciences,
     Univ. de l'Etat Mons, Av. Maistriau 19, B-7000 Mons, Belgium
    \label{AIM}}
\titlefoot{Physics Laboratory, University of Athens, Solonos Str.
     104, GR-10680 Athens, Greece
    \label{ATHENS}}
\titlefoot{Department of Physics, University of Bergen,
     All\'egaten 55, NO-5007 Bergen, Norway
    \label{BERGEN}}
\titlefoot{Dipartimento di Fisica, Universit\`a di Bologna and INFN,
     Via Irnerio 46, IT-40126 Bologna, Italy
    \label{BOLOGNA}}
\titlefoot{Centro Brasileiro de Pesquisas F\'{\i}sicas, rua Xavier Sigaud 150,
     BR-22290 Rio de Janeiro, Brazil \\
     \indent~~and Depto. de F\'{\i}sica, Pont. Univ. Cat\'olica,
     C.P. 38071 BR-22453 Rio de Janeiro, Brazil \\
     \indent~~and Inst. de F\'{\i}sica, Univ. Estadual do Rio de Janeiro,
     rua S\~{a}o Francisco Xavier 524, Rio de Janeiro, Brazil
    \label{BRASIL}}
\titlefoot{Coll\`ege de France, Lab. de Physique Corpusculaire, IN2P3-CNRS,
     FR-75231 Paris Cedex 05, France
    \label{CDF}}
\titlefoot{CERN, CH-1211 Geneva 23, Switzerland
    \label{CERN}}
\titlefoot{Institut de Recherches Subatomiques, IN2P3 - CNRS/ULP - BP20,
     FR-67037 Strasbourg Cedex, France
    \label{CRN}}
\titlefoot{Now at DESY-Zeuthen, Platanenallee 6, D-15735 Zeuthen, Germany
    \label{DESY}}
\titlefoot{Institute of Nuclear Physics, N.C.S.R. Demokritos,
     P.O. Box 60228, GR-15310 Athens, Greece
    \label{DEMOKRITOS}}
\titlefoot{FZU, Inst. of Phys. of the C.A.S. High Energy Physics Division,
     Na Slovance 2, CZ-180 40, Praha 8, Czech Republic
    \label{FZU}}
\titlefoot{Dipartimento di Fisica, Universit\`a di Genova and INFN,
     Via Dodecaneso 33, IT-16146 Genova, Italy
    \label{GENOVA}}
\titlefoot{Institut des Sciences Nucl\'eaires, IN2P3-CNRS, Universit\'e
     de Grenoble 1, FR-38026 Grenoble Cedex, France
    \label{GRENOBLE}}
\titlefoot{Helsinki Institute of Physics, P.O. Box 64,
     FIN-00014 University of Helsinki, Finland
    \label{HELSINKI}}
\titlefoot{Joint Institute for Nuclear Research, Dubna, Head Post
     Office, P.O. Box 79, RU-101 000 Moscow, Russian Federation
    \label{JINR}}
\titlefoot{Institut f\"ur Experimentelle Kernphysik,
     Universit\"at Karlsruhe, Postfach 6980, DE-76128 Karlsruhe,
     Germany
    \label{KARLSRUHE}}
\titlefoot{Institute of Nuclear Physics,Ul. Kawiory 26a,
     PL-30055 Krakow, Poland
    \label{KRAKOW1}}
\titlefoot{Faculty of Physics and Nuclear Techniques, University of Mining
     and Metallurgy, PL-30055 Krakow, Poland
    \label{KRAKOW2}}
\titlefoot{Universit\'e de Paris-Sud, Lab. de l'Acc\'el\'erateur
     Lin\'eaire, IN2P3-CNRS, B\^{a}t. 200, FR-91405 Orsay Cedex, France
    \label{LAL}}
\titlefoot{School of Physics and Chemistry, University of Lancaster,
     Lancaster LA1 4YB, UK
    \label{LANCASTER}}
\titlefoot{LIP, IST, FCUL - Av. Elias Garcia, 14-$1^{o}$,
     PT-1000 Lisboa Codex, Portugal
    \label{LIP}}
\titlefoot{Department of Physics, University of Liverpool, P.O.
     Box 147, Liverpool L69 3BX, UK
    \label{LIVERPOOL}}
\titlefoot{Dept. of Physics and Astronomy, Kelvin Building,
     University of Glasgow, Glasgow G12 8QQ
    \label{GLASGOW}}
\titlefoot{LPNHE, IN2P3-CNRS, Univ.~Paris VI et VII, Tour 33 (RdC),
     4 place Jussieu, FR-75252 Paris Cedex 05, France
    \label{LPNHE}}
\titlefoot{Department of Physics, University of Lund,
     S\"olvegatan 14, SE-223 63 Lund, Sweden
    \label{LUND}}
\titlefoot{Universit\'e Claude Bernard de Lyon, IPNL, IN2P3-CNRS,
     FR-69622 Villeurbanne Cedex, France
    \label{LYON}}
\titlefoot{Dipartimento di Fisica, Universit\`a di Milano and INFN-MILANO,
     Via Celoria 16, IT-20133 Milan, Italy
    \label{MILANO}}
\titlefoot{Dipartimento di Fisica, Univ. di Milano-Bicocca and
     INFN-MILANO, Piazza della Scienza 2, IT-20126 Milan, Italy
    \label{MILANO2}}
\titlefoot{IPNP of MFF, Charles Univ., Areal MFF,
     V Holesovickach 2, CZ-180 00, Praha 8, Czech Republic
    \label{NC}}
\titlefoot{NIKHEF, Postbus 41882, NL-1009 DB
     Amsterdam, The Netherlands
    \label{NIKHEF}}
\titlefoot{National Technical University, Physics Department,
     Zografou Campus, GR-15773 Athens, Greece
    \label{NTU-ATHENS}}
\titlefoot{Physics Department, University of Oslo, Blindern,
     NO-0316 Oslo, Norway
    \label{OSLO}}
\titlefoot{Dpto. Fisica, Univ. Oviedo, Avda. Calvo Sotelo
     s/n, ES-33007 Oviedo, Spain
    \label{OVIEDO}}
\titlefoot{Department of Physics, University of Oxford,
     Keble Road, Oxford OX1 3RH, UK
    \label{OXFORD}}
\titlefoot{Dipartimento di Fisica, Universit\`a di Padova and
     INFN, Via Marzolo 8, IT-35131 Padua, Italy
    \label{PADOVA}}
\titlefoot{Rutherford Appleton Laboratory, Chilton, Didcot
     OX11 OQX, UK
    \label{RAL}}
\titlefoot{Dipartimento di Fisica, Universit\`a di Roma II and
     INFN, Tor Vergata, IT-00173 Rome, Italy
    \label{ROMA2}}
\titlefoot{Dipartimento di Fisica, Universit\`a di Roma III and
     INFN, Via della Vasca Navale 84, IT-00146 Rome, Italy
    \label{ROMA3}}
\titlefoot{DAPNIA/Service de Physique des Particules,
     CEA-Saclay, FR-91191 Gif-sur-Yvette Cedex, France
    \label{SACLAY}}
\titlefoot{Instituto de Fisica de Cantabria (CSIC-UC), Avda.
     los Castros s/n, ES-39006 Santander, Spain
    \label{SANTANDER}}
\titlefoot{Inst. for High Energy Physics, Serpukov
     P.O. Box 35, Protvino, (Moscow Region), Russian Federation
    \label{SERPUKHOV}}
\titlefoot{J. Stefan Institute, Jamova 39, SI-1000 Ljubljana, Slovenia
     and Laboratory for Astroparticle Physics,\\
     \indent~~Nova Gorica Polytechnic, Kostanjeviska 16a, SI-5000 Nova Gorica, Slovenia, \\
     \indent~~and Department of Physics, University of Ljubljana,
     SI-1000 Ljubljana, Slovenia
    \label{SLOVENIJA}}
\titlefoot{Fysikum, Stockholm University,
     Box 6730, SE-113 85 Stockholm, Sweden
    \label{STOCKHOLM}}
\titlefoot{Dipartimento di Fisica Sperimentale, Universit\`a di
     Torino and INFN, Via P. Giuria 1, IT-10125 Turin, Italy
    \label{TORINO}}
\titlefoot{INFN,Sezione di Torino, and Dipartimento di Fisica Teorica,
     Universit\`a di Torino, Via P. Giuria 1,\\
     \indent~~IT-10125 Turin, Italy
    \label{TORINOTH}}
\titlefoot{Dipartimento di Fisica, Universit\`a di Trieste and
     INFN, Via A. Valerio 2, IT-34127 Trieste, Italy \\
     \indent~~and Istituto di Fisica, Universit\`a di Udine,
     IT-33100 Udine, Italy
    \label{TU}}
\titlefoot{Univ. Federal do Rio de Janeiro, C.P. 68528
     Cidade Univ., Ilha do Fund\~ao
     BR-21945-970 Rio de Janeiro, Brazil
    \label{UFRJ}}
\titlefoot{Department of Radiation Sciences, University of
     Uppsala, P.O. Box 535, SE-751 21 Uppsala, Sweden
    \label{UPPSALA}}
\titlefoot{IFIC, Valencia-CSIC, and D.F.A.M.N., U. de Valencia,
     Avda. Dr. Moliner 50, ES-46100 Burjassot (Valencia), Spain
    \label{VALENCIA}}
\titlefoot{Institut f\"ur Hochenergiephysik, \"Osterr. Akad.
     d. Wissensch., Nikolsdorfergasse 18, AT-1050 Vienna, Austria
    \label{VIENNA}}
\titlefoot{Inst. Nuclear Studies and University of Warsaw, Ul.
     Hoza 69, PL-00681 Warsaw, Poland
    \label{WARSZAWA}}
\titlefoot{Fachbereich Physik, University of Wuppertal, Postfach
     100 127, DE-42097 Wuppertal, Germany \\
\noindent
{$^\dagger$~deceased}
    \label{WUPPERTAL}}
\addtolength{\textheight}{-10mm}
\addtolength{\footskip}{5mm}
\clearpage
\headsep 30.0pt
\end{titlepage}
%
\pagenumbering{arabic} 
\setcounter{footnote}{0} %
\large
\section{Introduction}
The inclusive production of mesons has been a subject of long-standing study 
at LEPI, as it provides insight into the nature of 
fragmentation of quarks and gluons into hadrons. So far studies have been 
done on the $S$-wave mesons (both $^1S_0$ and $^3S_1$) such as $\pi$ and 
$\rho$, as well as certain $P$-wave mesons $f_2(1270)$, $K^*_2(1430)$ and
$f'_2(1525)$ (i.e. $^3P_2$) and $f_0(980)$ and $a_0(980)$ 
(i.e. $^3P_0$)~\cite{delph3,opal,PDG}. Very little is known
about the production of mesons belonging to other $P$-wave multiplets (i.e.
$^3P_1$ and $^1P_1$).  For the first time, we present in this paper a study 
of the inclusive production of two $J^{PC}=1^{++}$ mesons, the $f_1(1285)$ 
and the $f_1(1420)$ (i.e. $^3P_1$).

There are at least four known nonstrange isoscalar mesons~\cite{PDG},
$I^G(J^{PC})=0^+(1^{++})$ and $I^G(J^{PC})=0^+(0^{-+})$, in the mass region 
between 1.2 and 1.6 \gevcc, which couple to the decay channel 
$(K\bar K\pi)^0$. These are the $f_1(1285)$, $\eta(1295)$, $f_1(1420)$ and 
$\eta(1440)$. All are seen prominently in the peripheral production from
$\pi^-p$ interactions ~\cite{PDG}, indicating that, despite their decay into
$(K\bar K\pi)^0$, they are mostly $n\bar n$ states, where $n=\{u,d\}$. There 
exist possibly two additional states, $I^G(J^{PC})=?^-(1^{+-})$ $h_1(1380)$ 
and $I^G(J^{PC})=0^+(1^{++})$ $f_1(1510)$, which may harbour a large $s\bar s$
content, as they are produced with considerable cross-sections in the 
peripheral reactions involving $K^-p$ interactions~\cite{PDG}. Given this 
complexity in the $(K\bar K\pi)^0$ systems, it is important to find which 
resonances among these are readily excited in inclusive hadronic $Z$ decays.

The DELPHI data for this study are based on the neutral $K\bar K\pi$ channel 
in the reaction \bln{rc0}  Z\to (K_{_S}K^{\pm}\pi^{\mp})+X^0 
\eln
Section 2 is devoted to the selection process for the event sample collected 
for this analysis. The $K\bar K\pi$ mass spectra are studied in Section 3. It is 
shown that the selection of the events with low $M(K_{_S}\,K^\pm)$ mass is the 
crucial criterion to reveal the presence of two signals in the $f_1(1285)$ and 
$f_1(1420)$ mass regions. A partial-wave analysis, carried out to explore 
the spin-parity content of the two signals, is described in Section 4. 
The measurement of the production rates and differential cross-sections
is presented in Section 5. Conclusions are given in Section 6.

\section{Experimental Procedure}
The analysis presented here is based on a data sample of 3.4 million hadronic 
$Z$ decays collected from 1992 to 1995 with the DELPHI detector at LEP. A 
detailed description of the DELPHI detector and its performance can be found 
elsewhere~\cite{delph1,delph2}.

  The charged particle tracks have been measured in the 1.2 T magnetic field 
by a set of tracking detectors. The average momentum resolution for charged
particles in hadronic final states, $\Delta (1/p)$, is usually between 0.001 
and 0.01, depending on their momentum as well as on which detectors are included
in the track fit.

  A charged particle has been accepted in this analysis if its momentum $p$ 
is greater than 100 \mevc, its momentum error $\Delta p/p$ is less than 1 
and its impact parameter with respect to the nominal crossing point is within 
4 cm in the transverse ($xy$) plane and 4 cm/$\sin{\theta}$ along the beam 
direction ($z$-axis), $\theta$ being the polar angle of the track.

  Hadronic events are then selected by requiring at least 5 charged 
particles, 3 GeV as minimum energy of the charged particles in 
each hemisphere of the event (defined with respect to the beam direction) and 
total energy of the charged particles of at least 12\% of the centre-of-mass 
energy. 
The contamination from events due to beam-gas scattering and to 
$\gamma$-$\gamma$ interactions is estimated to be less than 0.1\% and the 
background from $\tau^+\,\tau^-$ events is less than 0.2\% of the 
accepted events.

  After the event selection, in order to ensure a better signal-to-background 
ratio for the resonances in the $K_{_S} K^\pm\pi^\mp$ invariant mass system, 
tighter requirements have been imposed on the track impact parameters with 
respect to the nominal crossing point, i.e. they have to be within 0.2 cm in 
the transverse plane and 0.4 cm/$\sin{\theta}$ along the beam direction.

  $K^\pm$ identification has been provided by the Barrel Ring Imaging Cherenkov
(BRICH)
detector for particles with momenta above 700 \mevc, while the ionization loss
measured in the Time Projection Chamber (TPC) has been used for momenta above 
100 \mevc. The corresponding identification tags are based on the combined 
probabilities, derived from the average Cherenkov angle and the number of 
observed photons in the BRICH detector, as well as the measured 
${\rm d}E/{\rm d}x$ in the TPC. Cuts on the tags have been applied to achieve 
the best signal-to-background ratio, while rejecting $e^\pm$, $\mu^\pm$,
$p$ and $\bar p$ tracks. A more detailed description of the identification 
tags can be found in Ref.\,\cite{ident1}. In the present case, the $K^\pm$ 
identification efficiency (typically 50\% over the kaon momentum range of 
this analysis \cite{ident1}) 
has been estimated by comparing the $\phi\ (1020)$ to $K^+K^-$ signal in the 
experimental data with a sample of simulated events generated with 
JETSET~\cite{pyth4}
tuned with the DELPHI parameters~\cite{delph5} and passed through the detector 
simulation program DELSIM~\cite{delph2}. Agreement within $\pm$4\% is observed 
between the data and the simulation.

  The $K_{_S}$ candidates are detected by their decay in flight into 
$\pi^+\pi^-$. The details of the reconstruction method and the various cuts 
applied are described in Ref.\,\cite{delph7}. Our selection process consists 
of taking the $V^0$'s passing the standard criteria for quality of the 
reconstruction plus a mass cut given by 0.45 $< M(\pi^+\pi^-) <$ 0.55 \gevcc.

   After all the above cuts, only events with at least one $K_{_S}K^+\pi^-$ 
or $K_{_S}K^-\pi^+$ combination have been kept in the present analysis, 
corresponding to a sample of 547k events.

\vskip12mm

\section{$K_{_S}K^{\pm}\pi^{\mp}$ Mass Spectra}

\begin{figure}[htb]\vskip-10mm
\begin{center}\mbox{
\epsfig{file=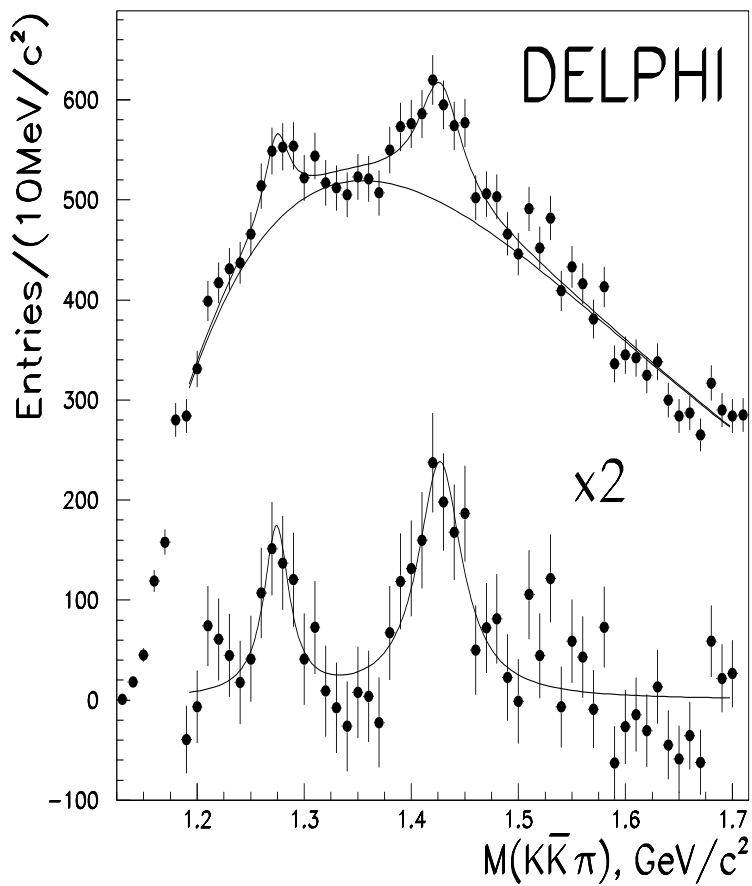,height=15.0cm,width=12.0cm}
}\end{center}
\vskip-13mm
\caption{Invariant mass distributions for the system $K_{_S}\,K^\pm\,\pi^\mp$ 
with a mass cut $M(K_{_S}\,K^\pm)<1.04$ \gevcc.  The two solid curves in the 
upper part of the histogram describe Breit-Wigner fits over a smooth
background (see text).  The lower histogram and the solid
curve give the same fits with the background subtracted and amplified
by a factor of two.}
\label{fg02}
\end{figure}

The $K_{_S}\,K^\pm\,\pi^\mp$ invariant mass distribution
is shown in Fig.\,\ref{fg01}a).  Also shown in the figure is the same
mass spectrum with a $K^*(892)$ selection ($0.822<M(K\pi)<0.962$ \gevcc), 
which would be appropriate
if the decay of a resonance had proceeded through a $K^*(892)$
intermediate state.  Neither histogram shows a
visible enhancement in the mass region
between 1.2 to 1.6 \gevcc. This is due to the enormous background
in this mass region coming from the high number of $K_{_S}K^{\pm}\pi^{\mp}$ 
combinations per event (11 on average) in inclusive $Z$ decays.  The key 
to a successful study of the $f_1(1285)$ and $f_1(1420)$ under the
circumstances is to select events with low $(K_{_S}\,K^\pm)$ mass
(Fig.\,\ref{fg01}b)). This has
the effect of selecting both the possible $a_0(980)^\pm\pi^\mp$ decay mode 
and, in case of $K^*(892)\bar K+c.c.$ decay,
the interference region of the two $K^*(892)$ bands on the decay Dalitz plot,
while reducing substantially the general background for the $K\bar K\pi$ 
system.
Varying this cut on the Monte Carlo generated events suggests
a mass cut $M(K_{_S}\,K^\pm)\leq1.04$ \gevcc\ to maximize both 
$f_1(1285)$ and $f_1(1420)$ signals over background. The application of
this cut on the experimental data is shown in Fig.\,\ref{fg02}, where two 
clear peaks are now seen in these mass regions where
the mass resolution is 8 and 9 \mevcc\ respectively.
Based on the Monte Carlo 
generated event sample, we have verified that neither signal was 
a reflection of resonances whose mass is in the 1.0 to 1.5 \gevcc\ range,
such as the $\phi\ (1020)$ and the $K_1(1270)$ mesons nor was faked by a 
possible misidentification of kaons or pions coming from the decay of
these resonances.

\begin{minipage}[b]{8cm}
\includegraphics[width=7.5cm]{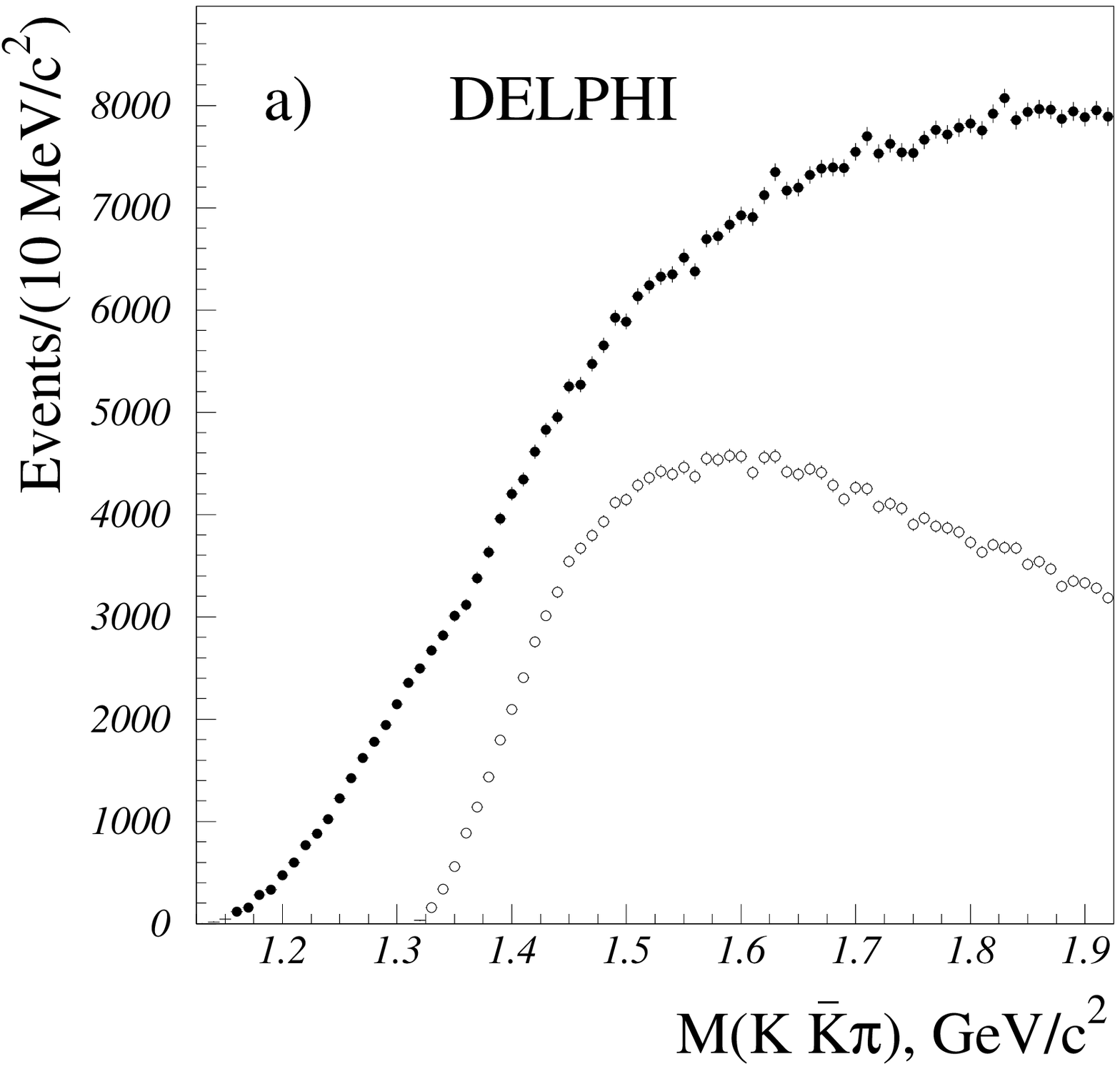}
\end{minipage}
\begin{minipage}[b]{8cm}
\includegraphics[width=7.5cm]{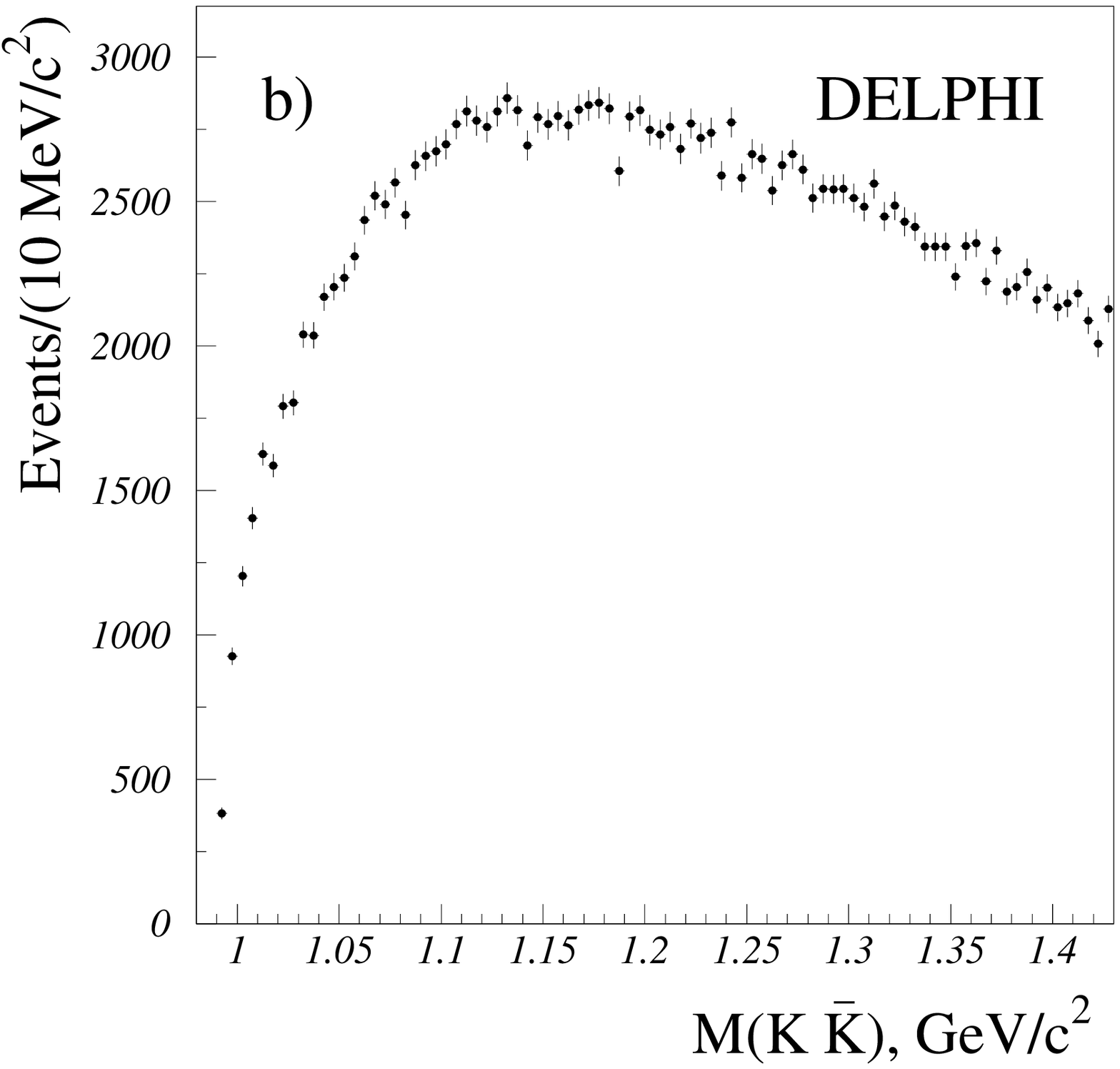}
\end{minipage}

\begin{figure}
\caption{$K_{_S}\,K^\pm\,\pi^\mp$ a) 
and $K_{_S}\,K^\pm$ b) invariant mass distributions from the 
$Z$ decays with the DELPHI detector at LEP I.  The histograms with solid 
circles are for the full data sample, that one with open circles is for data 
with a $0.822<M(K\pi)<0.962$ \gevcc\ - $K^*$ selection.
  }   
\label{fg01}
\end{figure}


\bigskip

To estimate the background under the signals, we have used the Monte Carlo 
generated event sample from which we have removed
all mesons with a major decay mode into $(K\bar K\pi)^0$ in the mass 
region 1.2 to 1.6 \gevcc.
The resulting $(K\bar K\pi)^0$ mass spectrum was fitted 
between 1.15 and 1.7 \gevcc\ with a background function
\bln{s3a}
   f_b(M)=(M-M_0)^{\textstyle\alpha_1}\,
                \exp(\alpha_2\,M+\alpha_3\,M^2),
\eln  
where $M$ and $M_0$ are the effective masses of the $(K\bar K\pi)^0$ 
system and its threshold, respectively, and $\alpha_i$ are the 
fitted parameters ($\alpha_1=0.7\pm0.1$, $\alpha_2=5.8\pm2.0$, 
$\alpha_3=-3.1\pm0.6$). Then we have fitted the experimental 
$(K\bar K\pi)^0$ spectrum from 1.19 to 1.7 \gevcc\
with the background function $f_b(M)$ determined above, adding two 
$S$-wave Breit-Wigner forms 
\bln{s3b}
   f_r(M)={\Gamma_r^2\over(M-M_r)^2+(\Gamma_r/2)^2}
\eln
We have not used the relativistic angular-momentum dependent 
Breit-Wigner form in (3) because such a form would require
a complete knowledge of the quantum numbers of the resonances 
as well as the branching ratios for all possible decay modes and
the corresponding orbital angular momenta. The limited statistics, 
the high background under the signals are additional reasons to adopt the
simple Breit-Wigner form (3).
The fitted masses and widths, $M_r$ and $\Gamma_r$, given in Table 1, 
are compatible with the PDG ~\cite{PDG} values for the 
$f_1(1285)$ and $f_1(1420)$ resonances. The
systematic uncertainty on the masses has been estimated from the various
background fits (described later) to be about 1--2 \mevcc\ and has been
added quadratically to the statistical error.
It should be noted that 
the parameters of the first peak are not compatible with those
of the $\eta(1295)$.
\vskip6mm
\noindent
\def\arraystretch{1.5}
\begin{center}
\begin{minipage}[]{100mm}
\tbhead{1}{Fitted parameters and numbers of events}
\begin{tabular}{|r@{$\,\pm\,$}l@{\hspace{12pt}}|
                     r@{$\,\pm\,$}l@{\hspace{12pt}}|
                                     r@{$\,\pm\,$}l@{\hspace{12pt}}|}
\hline\hline
\multicolumn{2}{|c|}{Mass (\mevcc)}&
\multicolumn{2}{c|}{Width (\mevcc)}&
\multicolumn{2}{c|}{Events}\\
\hline
\hspace{8pt}1274&6 & \hspace{16pt}29&12 
        & \hspace{8pt}358&$93\,({\rm stat})\pm59\,({\rm sys})$\\
\hspace{8pt}1426&6 & \hspace{16pt}51&14 
        & \hspace{8pt}870&$128\,({\rm stat})\pm136\,({\rm sys})$\\
\hline\hline
\end{tabular}
\end{minipage}
\end{center}
\def\arraystretch{1.0}
\noindent
\vskip3mm
 The numbers of events in Table 1
correspond to the fit shown in Fig.~\ref{fg02}, 
where the widths of the two peaks have been fixed to the
fitted values, while the background parameters were left free.
\vskip-20mm

    The main sources of systematic uncertainty come from the various cuts and
selection criteria applied for the $V^0$ reconstruction and the charged $K$ 
identification on the one hand and 
the conditions of the fit procedure on the other. 
To estimate the first type of error, we have compared 
the $K_{_S}K^\pm$ mass distributions 
of the simulated sample with the real data. 
Normalized to the same number of events, 
the distributions agree within $\pm$7\%, in the low $K_{_S}K^\pm$ mass region.

   The $f_1(1285)$ and $f_1(1420)$ signals show up
over a large background ($\sim80$\%). Variations of the background shape and 
amplitude induce sizable fluctuations of the fitted numbers of signal events. 
To quantify this effect, we have performed various series of fits, one varying 
the mass range of the fit, another leaving free the background parameters while
fixing the width of the signals, another with a polynomial shape for the 
background, thereby allowing the background level and shape to fluctuate. In 
this way we estimate 
the uncertainty of the number of fitted events to be $\pm$15\% for the 
$f_1(1285)$ and $\pm$14\% for the $f_1(1420)$. The systematic uncertainties 
have been added quadratically and are shown in Table 1.

   The overall efficiencies for the two states have been estimated from
the Monte Carlo simulated events to be:
\begin{eqnarray}
   (0.063&  \pm & 0.003)\%\quad{\rm for}\quad f_1(1285),\nonumber\\
   (0.45& \pm & 0.02)\%\quad{\rm for}\quad f_1(1420).
\end{eqnarray}
The quoted numbers include the following corrections for the
$f_1(1285)$ and $f_1(1420)$ respectively: branching ratios to $K\bar K\pi$
$(0.09, 1.)$, fractions of final states with charged pion  
$(1/2, 2/3)$, branching ratio of $K^0 \to \pi^+\pi^- = 1/2 \times0.686 = 0.343
$, reconstruction and identification efficiency for the selected
events $(0.058, 0.061)$ and correction factor $(0.70, 0.32)$ 
for the $M(K_{_S}\,K^\pm)\leq1.04$ \gevcc\ mass cut. The quoted errors
are statistical errors on the Monte Carlo sample. 

\section{Partial-Wave Analysis}
In an attempt to get more information on the spin-parity content of the two
signals we have performed a mass-dependent partial-wave analysis (PWA) of the 
$K_{_S}K^{\pm}\pi^{\mp}$ system.
There have been many 3-body partial-wave analyses; the reader
may consult PDG~\cite{PDG} for earlier references, for example, on
$a_1(1260)$, $a_2(1320)$, $K_1(1270/1400)$ or $K_2(1770)$.  For the first
time, we apply the same technique to a study of the $(K\bar K\pi)^0$
system from the inclusive decay of the $Z$ at LEP.

  A spin-parity analysis of the system composed of three
pseudoscalars requires five variables,
which may be chosen to be the three Euler angles defining the orientation
of the 3-body system in its suitably-chosen rest frame and
two effective masses describing the decay Dalitz plot.  We have chosen to
employ the so-called Dalitz plot analysis, integrating over the three
Euler angles. This entails an essential simplification in the number of
parameters required in the analysis, as the decay amplitudes involving the
$D$-functions defined over the three Euler angles and their appropriate
decay-coupling constants, are orthogonal for different spins and 
parities~\cite{ch0}. In these conditions, the mass-dependent PWA comes 
to fitting Dalitz plots, thus providing the contribution of the various 
$J^{PC}$ waves as a function of the $M(K\bar K\pi)$.
The actual fitting of the data is done by using the maximum-likelihood method,
in which the normalization integrals are evaluated with the Monte Carlo 
events~\cite{e852a}, thus taking into account the 
$M(K_{_S}\,K^\pm)\leq1.04$ \gevcc\ cut. The comparison between fits is made
on the basis of their maximum likelihood values and their description of the 
$(K\bar K\pi)$, $(K\pi)$ and $(K\bar K)$ mass distributions.

    The first step of the analysis is to parametrize the background under
the two signals.
This background accounts for different processes
with, for example, different overall multiplicities. From a study of the side
bands (away from the resonance regions), it has been determined 
that the background contains substantial amounts of $a_0(980)$ and $K^*(892)$ 
unassociated with the resonances.  We have thus assumed that the background, 
which should not interfere with the signals, can be described by a 
non-interfering superposition of a constant three-body phase-space 
term and the partial waves $I^G(J^{PC})=0^+(0^{-+})\,a_0(980)\pi$ ($S$-wave),
$0^+(1^{++})\,(K^*(892)\bar K+c.c.)$ ($S$-wave) 
and $0^-(1^{+-})\,(K^*(892)\bar K+c.c.)$ ($S$-wave).
The $(K\bar K\pi)$ mass dependence of the background components is 
parametrized by the phase-space-like form given by (2), but with $\alpha_3=0$. 

The signals themselves are represented by a set of quasi two-body amplitudes 
which have the form of Breit-Wigner functions multiplied by 
spin-parity terms~\cite{e852a}. Those include the $I^G(J^{PC})=0^+(1^{++})$, 
$0^+(0^{-+})$ and $0^-(1^{+-})$ partial waves, where the possible decay 
channels $a_0(980)\pi$ and $K^*(892)\bar K+c.c.$ are allowed to interfere 
within a given $J^{PC}$.  

The PWA was performed for $M(K\bar K\pi)$ in the $1.18\to1.66$ \gevcc\
mass range.
The first series aimed at determining the background contributions. 
 In this fit, the signals were assumed to be composed of the 
$\eta(1295)$ and $f_1(1285)$ for the first peak and for the second of 
the $\eta(1440)$, $h_1(1380)$ and $f_1(1420)$ resonances which
were parametrized as Breit-Wigner forms with masses and widths fixed to 
the PDG values~\cite{PDG}. The fit was checked to well reproduce the
$(K\pi)$ and $(K\bar K)$ mass distributions outside the regions of the
peaks.

 The following step consisted in fitting the spin-parity content of the
 two signals. For this the
individual background contributions were fixed to their fitted values, only 
the overall background rate was left free. The $J^{PC}$ amplitudes were
introduced individually to probe the spin-parity content of each signal with
the mass and width of the corresponding $(K\bar K\pi)^0$ resonance being fitted.

The results of these fits are the following. The 
$I^G(J^{PC})=0^+(1^{++})\,a_0(980)\pi$ and  $0^+(0^{-+})\,a_0(980)\pi$ waves
account equally well for the 1.28 \gevcc\ region, with the same maximum 
likelihood value, i.e the first peak is equally likely to be the $f_1(1285)$ 
or the $\eta(1295)$. On the other hand, if the mass and the width of the 
resonances are fixed to their PDG values, this region is better fitted by the 
$I^G(J^{PC})=0^+(1^{++})\,a_0(980)\pi$ wave. This reflects the fact that 
the first peak position is closer to the $f_1(1285)$ mass than to the
$\eta(1295)$ mass, as already noticed in the fit of the $M(K\bar K\pi)$ 
spectrum.

In the 1.4 \gevcc\ region, the maximum likelihood value 
is better for the $I^G(J^{PC})=0^+(1^{++})\,K^*(892)\bar K+c.c.$ wave over 
the $I^G(J^{PC})=0^+(0^{-+})$, the $0^-(1^{+-})\,K^*(892)\bar K+c.c.$ and 
the $0^+(1^{++})\,a_0(980)\pi$ waves by about 4, 8 and 9 units, respectively,
and thus favours the $f_1(1420)\to K^*\bar K$ hypothesis over the $\eta(1440)$ or
$h_1(1380)$ ones, as is verified on the projections of the individual fits
(not shown) on the $(K\bar K\pi)$, $(K\pi)$ and $(K\bar K)$ mass distributions. 

\begin{figure}[ht]\vskip-80mm
\begin{center}\mbox{
\epsfig{file=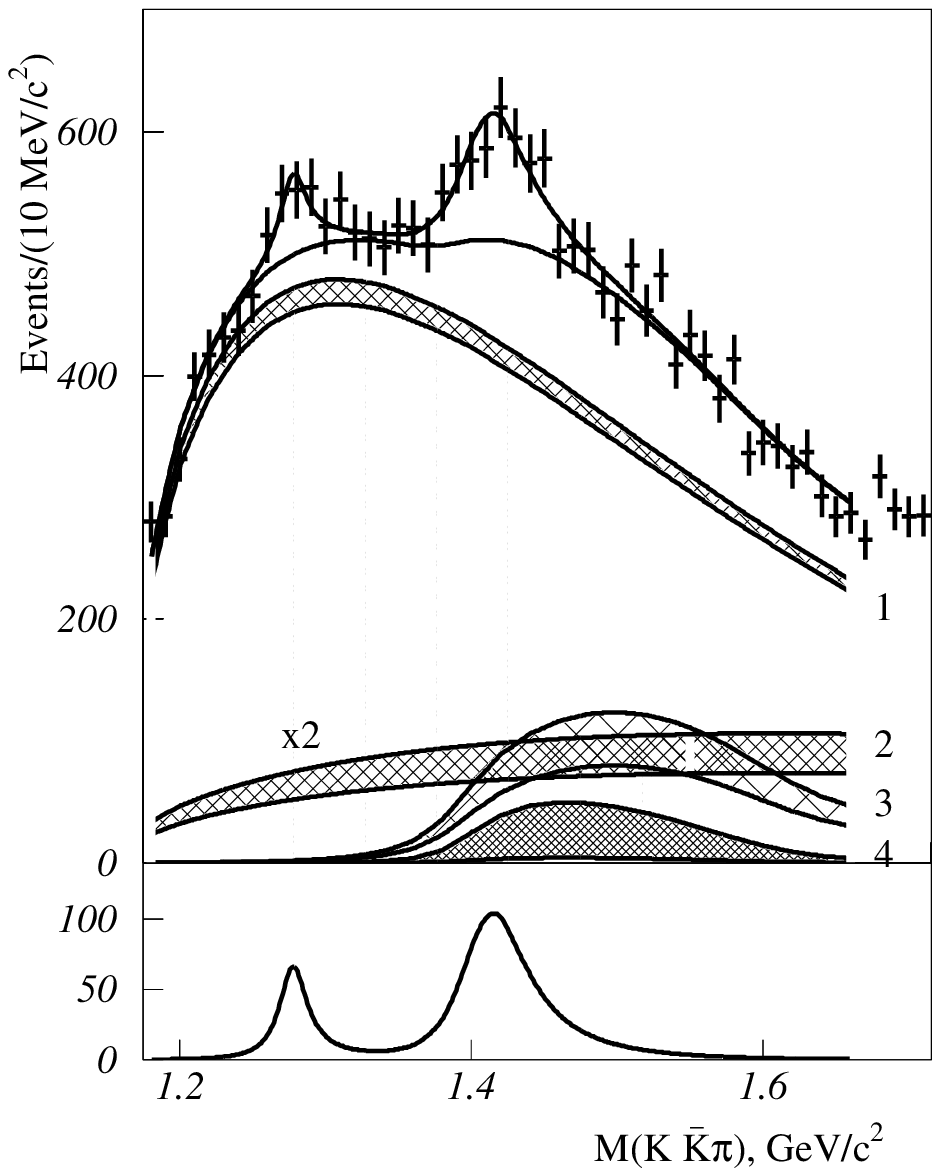,height=28.0cm,width=26.0cm}
}\end{center}
\vskip-77mm
\caption{$M(K_{_S}\,K^\pm\,\pi^\mp)$ distributions per 10 \mevcc\ with 
a breakdown into the partial waves for the signals in the lower histogram
and for the background shown as one error-band.
The signals consist of $1^{++}\,a_0(980)\pi$ for the first peak
and $1^{++}\,K^*(892)\bar K$ for the second peak.  The background is
composed of a non-interfering superposition of (1) isotropic phase-space 
distribution and the following partial waves:
(2) $0^{-+}\,a_0(980)\pi$, 
(3) $1^{++}\,K^*(892)\bar K$ and (4) $1^{+-}\,K^*(892)\bar K$ waves. 
The last three contributions are shown magnified by a factor of two.}
\label{fg03}
\end{figure}

The results of the best fit with the $I^G(J^{PC})=0^+(1^{++})\,a_0(980)\pi :
f_1(1285)$ and $0^+(1^{++})\,K^*(892)\bar K+c.c. : f_1(1420)$ amplitudes are 
shown in Fig.\,\ref{fg03} with the background contributions in the form of 
error-bands. The masses, the widths and the numbers of events found in the 
fit are statistically consistent with those given in Table 1. One observes
that the background events $I^G(J^{PC})= 0^+(1^{++})\,
(K^*(892)\bar K+c.c.)$ ($S$-wave) 
and $0^-(1^{+-})\,(K^*(892)\bar K+c.c.)$ ($S$-wave) exhibit a shape suggestive 
of a resonance. However this effect is simply the result of: i) our mass
dependent function which decreases rapidly at high $(K\bar K\pi)$ mass
to reproduce the fall off due to the $M(K_{_S}\,K^\pm)<1.04$ \gevcc\ cut;
ii) the $K^*(892)\bar K$ threshold around 1.4 \gevcc\  whose effect is gradual 
because of the finite width of the $K^*(892)$.

The major sources of systematic uncertainties come from the background
description and the conditions of the overall PWA fit. To estimate them we have
carried out series of fits leaving free the background contributions, 
then fixing the mass and width of the $(K\bar K\pi)^0$ resonances 
to their PDG values~\cite{PDG}. This was repeated with a polynomial background
in $M(K\bar K\pi)$ in place of the phase-space-like one. 
In all these fits, the partial waves were fitted concurrently, in the form of 
a non-interfering superposition, to estimate their relative contribution. 
All the fits confirm the observations made previously. 

Taking into account the systematic uncertainties computed from the various
fits, the number of $f_1(1285)$ and of $f_1(1420)$ events for the best 
fit of Fig.\,\ref{fg03} are $237\pm60{\rm (stat)}\pm70{\rm (syst)}$ and 
$711\pm100{\rm (stat)}\pm75{\rm (syst)}$, respectively, consistent within
systematic uncertainties with the 
values of the fit to the $M(K\bar K\pi)$ spectrum described in Section 3.

All fits confirm the dominance of the             
$I^G(J^{PC})=0^+(1^{++})\,K^*(892)\bar K+c.c.$ wave in the 1.4 \gevcc\ region.
The largest contributions of the $\eta(1440)$ and $h_1(1380)$, estimated from 
the highest fitted rates of the 
$I^G(J^{PC})=0^+(0^{-+})$ and $0^-(1^{+-})\,K^*(892)\bar K+c.c.$ waves,
correspond to production rates per hadronic $Z$ decay of 0.+0.007 and 
$(0.017^{+0.011}_{-0.015})$ respectively, assuming a $K^*(892)\bar K+c.c.$ 
branching ratio of 100\% for these resonances.


\section{Production Rates and Differential Cross-sections} 
From the histogram fit described in section 3.,
we have measured the production rate $\bra n\ket$ per hadronic $Z$ decay 
for $f_1(1285)$ and $f_1(1420)$. The results are
\begin{eqnarray}
   \bra n\ket&=&0.165\pm0.051\quad{\rm for}\quad f_1(1285),\nonumber\\
   \bra n\ket&=&0.056\pm0.012\quad{\rm for}\quad f_1(1420),
\end{eqnarray}
taking a $K\bar K\pi$ branching ratio of $(9.0\pm0.4)$\% for the $f_1(1285)$ 
and 100\% for the $f_1(1420)$~\cite{PDG}. 

\begin{figure}[htb]\vskip2mm
\begin{center}\mbox{
\epsfig{file=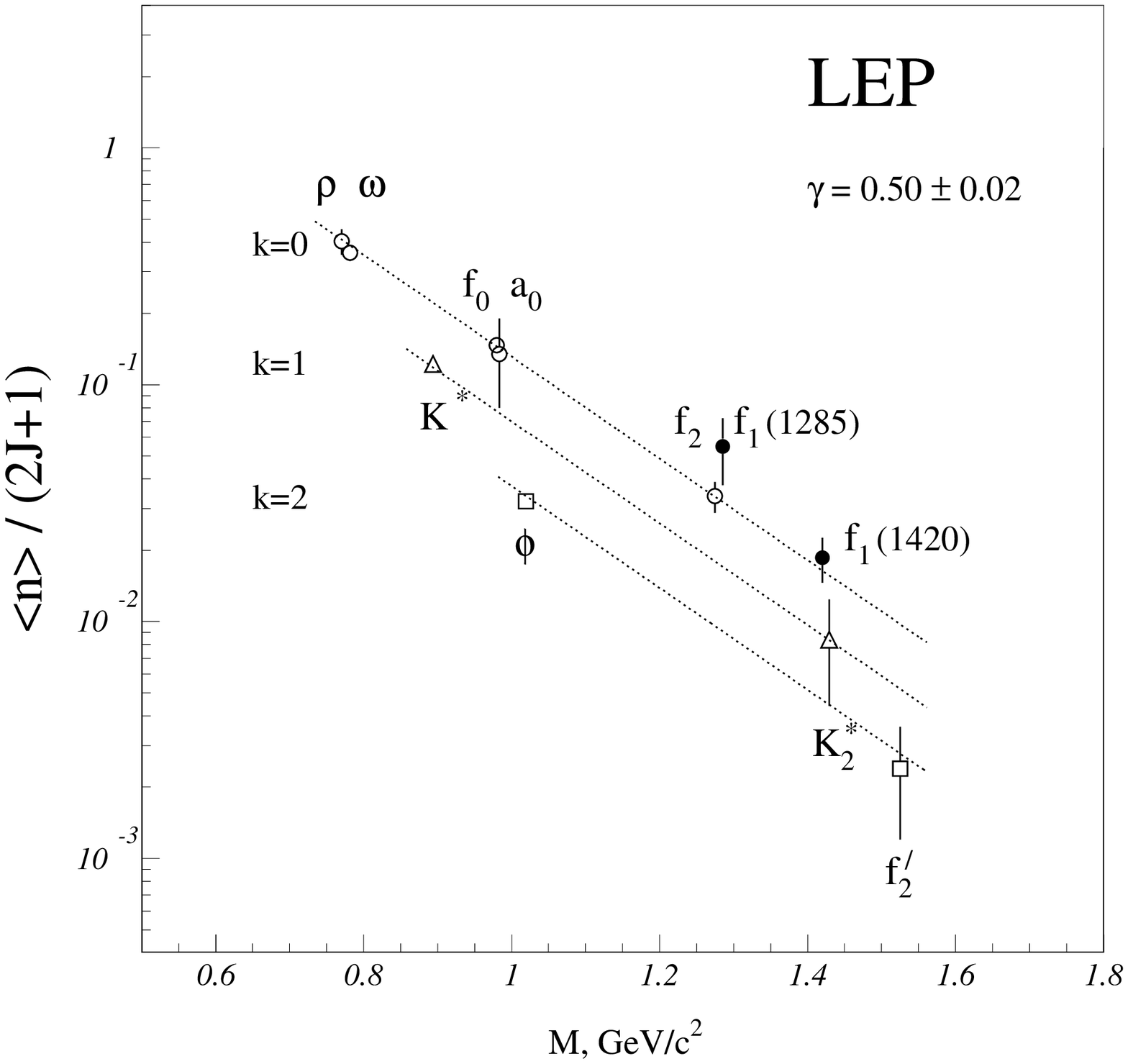,width=12.0cm}
}\end{center}
\caption{Total production rate per spin state and isospin for scalar, vector 
and tensor mesons as a function of the mass (open symbols) from ~\cite{uvr0}.
The two solid circles correspond to the $f_1(1285)$ and the $f_1(1420)$ 
measurements presented here.}
\label{fg04}
\end{figure}

The total production rates, per spin state and isospin, for the scalar, vector 
and tensor mesons with different strangeness, as a function of the 
mass~\cite{uvr0,chliap} are shown in Fig.\,\ref{fg04} for the averaged LEP 
data. To this figure we have added our measurements for comparison. It is 
seen that both $f_1(1285)$ and $f_1(1420)$ 
come close to the line corresponding to mesons whose constituents are thought 
to be of the type $n\bar n$. This suggests that both $f_1(1285)$ and 
$f_1(1420)$ have little $s\bar s$ content.
Indeed, the two states which are thought to be pure $s\bar s$
mesons, the $\phi$ and the $f'_2(1525)$, are down by a factor
$\gamma^k\approx1/4$ ($\gamma=0.50\pm0.02$ ~\cite{uvr0} and $k=2$, where 
k is the number of $s$ and $\bar s$ quarks in the meson), 
as shown in Fig.\,\ref{fg04}.  
A high strange quark content is highly unlikely given the production rate (5).

\begin{figure}[ht]
\begin{center}\mbox{
\epsfig{file=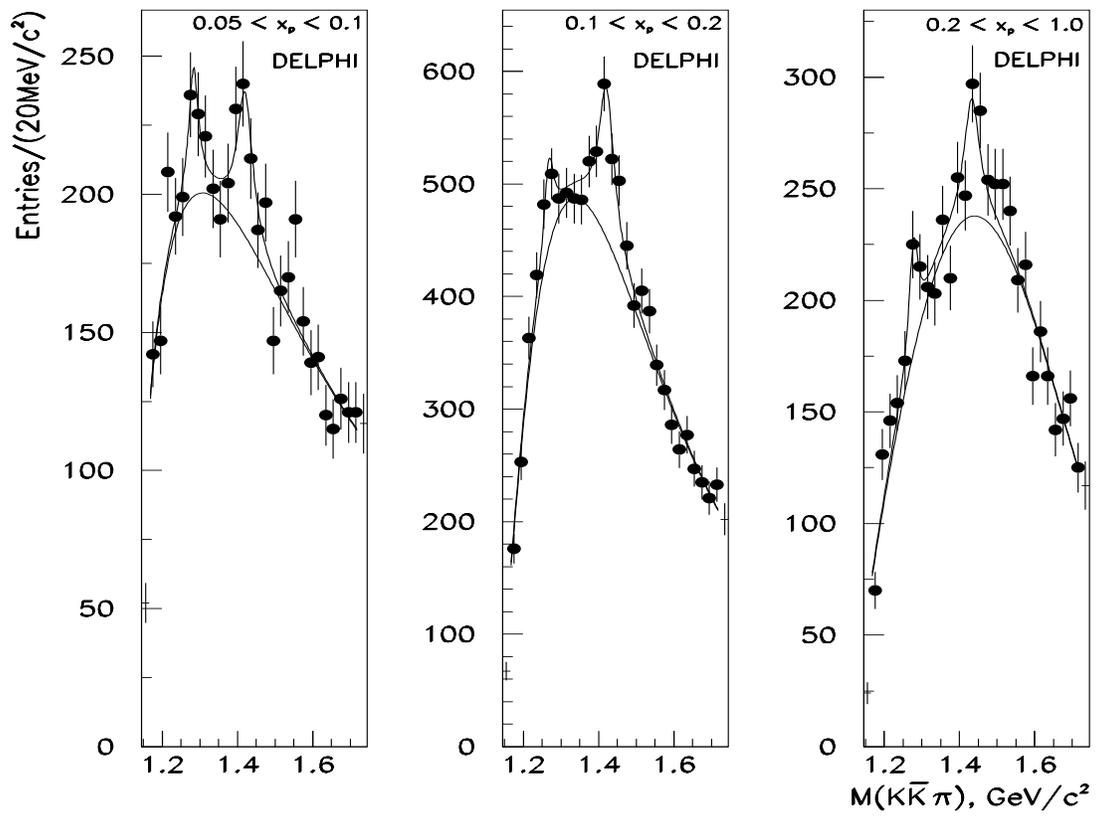,height=14.0cm,width=18.0cm}
}\end{center}
\caption{The $K_{_S}\,K^\pm\,\pi^\mp$ invariant mass spectra
for various $x_p$ intervals as indicated. Dots are the data, the solid lines 
show the result from the fit and the background contribution.}
\label{fg05}
\end{figure}

For completeness, we give in Fig.\,\ref{fg05} and in Table 2 the 
$f_1(1285)$ and $f_1(1420)$ differential rates and cross-sections
as a function of the scaled momentum $x_p$ 
($x_p$ = $p_{K\bar K\pi^0}$/$p_{\,\rm beam}$) for $x_p >.05$, as the signal to
background ratio is too small for lower momenta. Quantitative comparison with 
JETSET predictions is not possible in a meaningful way as there was no 
tuning for $f_1(1285)$ and $f_1(1420)$ and the implementation of the 
$(K\bar K\pi)^0$ decay of both resonances in JETSET had been done according to 
phase-space and not according to the correct spin-parity matrix element.
The small excess of events in Fig.\,\ref{fg05}, near $M(K_{_S}K^+\pi^-)= $ 1.55
\gevcc\, for $0.05< x_p <0.1$, is not significant enough 
for further study here.

\newpage
\vskip-16mm
\centerline{\parbox{136mm}{
\tbhead{2}{Measured production rates per hadronic event, differential
cross-sections and experimental efficiencies
for the $f_1(1285)$ and $f_1(1420)$, as functions of $x_p$.}
\begin{center}\def\arraystretch{1.4}
\begin{tabular}{|@{\hspace{8pt}}c@{\hspace{8pt}}|
   @{\hspace{8pt}}c@{\hspace{8pt}}|
   @{\hspace{8pt}}c@{\hspace{8pt}}|@{\hspace{8pt}}c@{\hspace{8pt}}|}
\hline\hline
$x_p$ interval & $f_1(1285)$ rate 
	& $(1/\sigma_h)(d\sigma/dx_p)$ & Efficiency\\
\hline
.05-.10 & 0.046$\pm$0.026 & 0.92$\pm$0.52 & $(6.5\pm0.7)\times10^{-4}$\\
.10-.20 & 0.053$\pm$0.024 & 0.53$\pm$0.24 & $(9.4\pm0.8)\times10^{-4}$\\
.20-1.0 & 0.051$\pm$0.022 & 0.06$\pm$0.03 & $(6.4\pm0.7)\times10^{-4}$\\
\hline
\end{tabular}\vskip16pt
\begin{tabular}{|@{\hspace{8pt}}c@{\hspace{8pt}}|
   @{\hspace{8pt}}c@{\hspace{8pt}}|
   @{\hspace{8pt}}c@{\hspace{8pt}}|@{\hspace{8pt}}c@{\hspace{8pt}}|}
\hline\hline
$x_p$ interval & $f_1(1420)$ rate 
	& $(1/\sigma_h)(d\sigma/dx_p)$ & Efficiency\\
\hline
.05-.10 & 0.018$\pm$0.006 & 0.36$\pm$0.12 & $(3.1\pm0.3)\times10^{-3}$\\
.10-.20 & 0.017$\pm$0.004 & 0.17$\pm$0.04 & $(8.5\pm0.5)\times10^{-3}$\\
.20-1.0 & 0.015$\pm$0.005 & 0.02$\pm$0.01 & $(3.9\pm0.3)\times10^{-3}$\\
\hline
\hline
\end{tabular}\def\arraystretch{1.0}
\end{center}
}}
\vfil\eject
\clearpage

\section{Conclusions}
  We have studied the inclusive production of two $(K\bar K\pi)^0$ states
in $Z$ decays at LEP I. The measured masses and widths are $1274\pm6$ 
and $29\pm12$ \mevcc\ for
the first peak and $1426\pm6$ and $51\pm14$ \mevcc\ for the second one,
compatible with those of the $f_1(1285)$ and $f_1(1420)$ mesons ~\cite{PDG}.
For the first time, a partial-wave analysis has been carried out
on the $(K\bar K\pi)^0$ system from the inclusive $Z$ decay.  While the 
results are ambiguous between
the $I^G(J^{PC})=0^+(1^{++})$ and $0^+(0^{-+})\,a_0(980)\pi$ waves in the 
1.28 \gevcc\ region, the second peak is uniquely consistent with 
$I^G(J^{PC})=0^+(1^{++})\,K^*(892)\bar K+c.c.$. On the other 
hand, the comparison of the hadronic production rate of these two states 
with a previous study of the production rate~\cite{uvr0,chliap} for the $S=1$ 
mesons (which included $^3S_1$, $^3P_0$ and $^3P_2$) suggests that their 
quantum numbers are very probably $I^G(J^{PC})=0^+(1^{++})$ and that their 
quark constituents are mainly of the type $n\bar n$, where $n=\{u,d\}$ and
thus confirms that these states are very likely the $f_1(1285)$ and 
$f_1(1420)$ mesons. Finally, we conclude that the mesons
$\eta(1295)$, $\eta(1440)$ and $h_1(1380)$ are less likely to be produced
in the inclusive $Z$ decays compared to the $f_1(1285)$ and $f_1(1420)$.

\subsection*{Acknowledgements}
\vskip 3 mm
 We are greatly indebted to our technical 
collaborators, to the members of the CERN-SL Division for the excellent 
performance of the LEP collider, and to the funding agencies for their
support in building and operating the DELPHI detector.\\
We acknowledge in particular the support of \\
Austrian Federal Ministry of Education, Science and Culture,
GZ 616.364/2-III/2a/98, \\
FNRS--FWO, Flanders Institute to encourage scientific and technological 
research in the industry (IWT), Federal Office for Scientific, Technical
and Cultural affairs (OSTC), Belgium,  \\
FINEP, CNPq, CAPES, FUJB and FAPERJ, Brazil, \\
Czech Ministry of Industry and Trade, GA CR 202/99/1362,\\
Commission of the European Communities (DG XII), \\
Direction des Sciences de la Mati$\grave{\mbox{\rm e}}$re, CEA, France, \\
Bundesministerium f$\ddot{\mbox{\rm u}}$r Bildung, Wissenschaft, Forschung 
und Technologie, Germany,\\
General Secretariat for Research and Technology, Greece, \\
National Science Foundation (NWO) and Foundation for Research on Matter (FOM),
The Netherlands, \\
Norwegian Research Council,  \\
State Committee for Scientific Research, Poland, SPUB-M/CERN/PO3/DZ296/2000,
SPUB-M/CERN/PO3/DZ297/2000 and 2P03B 104 19 and 2P03B 69 23(2002-2004)\\
JNICT--Junta Nacional de Investiga\c{c}\~{a}o Cient\'{\i}fica 
e Tecnol$\acute{\mbox{\rm o}}$gica, Portugal, \\
Vedecka grantova agentura MS SR, Slovakia, Nr. 95/5195/134, \\
Ministry of Science and Technology of the Republic of Slovenia, \\
CICYT, Spain, AEN99-0950 and AEN99-0761,  \\
The Swedish Natural Science Research Council,      \\
Particle Physics and Astronomy Research Council, UK, \\
Department of Energy, USA, DE-FG02-01ER41155, \\
EEC RTN contract HPRN-CT-00292-2002. \\
S.-U. Chung is grateful for the warm hospitality extended to him
by the CERN staff during his sabbatical year in the EP Division.
\brlist
\bibliographystyle{unsrt}
\brf{delph3} DELPHI Collab., P. Abreu {\it et al.},
             Phys. Lett. {\bf B449} (1999) 364.
\brf{opal}   OPAL Collab., R. Akers {\it et al.},
             Z. Phys. {\bf C68} (1995) 1.\\
	     K. Ackerstaff {\it et al.},
	     Eur Phys. J {\bf C4} (1998) 19;
	     ibid. {\bf C5} (1998) 411.	     
\brf{PDG}    Particle Data Group,
             K.Hagiwara et al., Phys. Rev. {\bf D66}, 010001 (2002).          
\brf{delph1} DELPHI Collab., P. Aarnio {\it et al.}, 
             Nucl. Inst. Meth. {\bf A303} (1991) 233.
\brf{delph2} DELPHI Collab., P. Abreu {\it et al.}, 
             Nucl. Inst. Meth. {\bf A378} (1996) 57.
\brf{ident1} DELPHI Collab., P. Abreu {\it et al.},
             Eur. Phys. J. {\bf C5}, (1998) 585.	     
\brf{pyth4}  T. Sj\"{o}strand, Comput. Phys. Comm. {\bf 82} (1994) 74.
\brf{delph5} DELPHI Collab., P. Abreu {\it et al.}, 
             Z. Phys. {\bf C73} (1996) 11.  
\brf{delph7} DELPHI Collab., P. Abreu {\it et al.}, 
             Z. Phys. {\bf C65} (1995) 587.
\brf{ch0}    See, for example, S. U. Chung, 
             `Spin Formalisms,' CERN preprint 71-8 (1971).
\brf{e852a}  E852 Collab., S. U. Chung {\it et al.}, 
             Phys. Rev. {\bf D60} (1999) 092001.
\brf{uvr0}   V. Uvarov, Phys. Lett. {\bf B511} (2001) 136.\\
             V. Uvarov, Phys. Lett. {\bf B482} (2000) 10.
\brf{chliap} P.V. Chliapnikov, Phys. Lett. {\bf B525} (2002) 1.
\erlist

\end{document}